\documentclass[12pt]{iopart}
\usepackage{iopams}
\usepackage{graphicx}
\usepackage[colorlinks,linkcolor=blue,anchorcolor=blue,citecolor=blue,urlcolor=blue]{hyperref}
\usepackage{bm}
\usepackage{color}
\expandafter\let\csname equation*\endcsname\relax
\expandafter\let\csname endequation*\endcsname\relax
\usepackage{amsmath}

\begin{document}

\title[Topological invariants in correlated Chern insulators]{Quantum cluster approach to the topological invariants in correlated Chern insulators}

\author{Zhao-Long Gu $^1$, Kai Li$^{1,2}$, and Jian-Xin Li$^{1,3}$}
\address{$^1$ National Laboratory of Solid State Microstructures and Department of Physics, Nanjing University, Nanjing 210093, China}
\address{$^2$ School of Physics and Engineering, Zhengzhou University, Zhengzhou 450001, China}
\address{$^3$ Collaborative Innovation Center of Advanced Microstructures, Nanjing University, Nanjing 210093, China}
\eads{\mailto{jxli@nju.edu.cn}}
\date{\today}

\begin{abstract}
\par We detect the topological properties of Chern insulators with strong Coulomb interactions by use of cluster perturbation theory and variational cluster approach. The common scheme in previous studies only involves the calculation of the interacting bulk Chern number within the natural unit cell by means of the so-called topological Hamiltonian. With close investigations on a prototype model, the half-filled Haldane Hubbard model, which is subject to both periodic and open boundary conditions, we uncover the unexpected failure of this scheme due to the explicit breaking of the translation symmetry. Instead, we assert that the faithful interacting bulk Chern number in the framework of quantum cluster approaches can be computed in the enlarged unit cell, which is free of the fault caused by the explicit translation symmetry breaking and consistent with the interacting bulk-edge correspondence.

\end{abstract}

\submitto{\NJP}
\maketitle

\section{Introduction}

\par Band insulators with nontrivial topology have become one of the central interests in condensed matter physics since the discovery of topological insulators \cite{HK_RMP2010,QZ_RMP2011}. Beyond the traditional Landau paradigm, topological insulators cannot be distinguished from ordinary ones by local order parameters, but are featured by gapless edges states on open boundaries \cite{H_PRL1988,K_PU2001,KM_PRL2005,BHZ_S2006,LYGJ_PRB2016,LGLW_NJP2017}. The existence of such edge modes is linked to the emergence of nonzero topological invariants of the bulk energy bands via the bulk-edge correspondence \cite{QWZ_PRB2006}. Thus the identification of a topological invariant is of particular importance to characterize different phases of band insulators. Indeed, in terms of such topological invariants constructed from the free Hamiltonians, a complete classification has been developed for non-interacting systems in different symmetry classes \cite{SRFL_PRB2008,CTSR_RMP2016}.

\par The presence of strong interactions between electrons enriches the physics of topological bands, and brings many challenges to this area at the same time \cite{HA_JPCM2013,R_RPP2018}. One of such demanding questions is how one can compute the topological invariant in an interacting system. A pioneering work by Volovik \cite{V2003} expresses the Chern number \cite{TKNN_PRL1982} of a two dimensional Chern insulator in terms of the single particle Green's functions. This formula is ready to be generalized to deal with correlated Chern insulators by a simple substitution of the free propagator with the interacting one. A great simplification of this formula was proposed in Ref. \cite{WZ_PRX2012} that the zero-energy single particle Green's function alone is sufficient to calculate the interacting Chern number via the Kubo formula. Later, the concept of the so-called topological Hamiltonian \cite{WY_JPCM2013}, i.e. the minus inverse of the dressed single particle Green's function, was shown to apply equally in other systems with diverse topological classifications \cite{WA_PRB2013,LEGW_PRB2013,HWGF_PRB2013,GMCBB_NJP2015,LS_PRB2018,SK_PRB2018}, such as the quantum spin Hall insulators \cite{LEGW_PRB2013,HWGF_PRB2013,GMCBB_NJP2015}, topological superconductors \cite{LS_PRB2018}, interacting Su-Schrieffer-Heeger chains \cite{SK_PRB2018}, etc.

\par Thus, the success of the topological Hamiltonian relies on a credible computation of the interacting single particle Green's functions. In recent years a variety of quantum many-body algorithms gathering under the name of quantum cluster approaches \cite{MJPH_RMP2005,S_arXiv2008} have been developed to calculate such single particle Green's functions of a broad range of strongly correlated fermionic systems \cite{SPP_PRL2000,HMJK_PRB2000,KSPB_PRL2001,PAD_PRL2003,ST_PRL2004,SS_PRL2008,BKSTP_EPL2009,KYXL_PRB2011,YXL_PRL2011,WRLH_PRB2012,YL_PRB2012,HD_PRL2013,LRTR_PRB2014,LYXL_PRB2014,RLRT_PRL2015} and bosonic systems \cite{KD_JPCM2006,KAL_PRB2011,YWDYL_PRB2018}. They share the basic idea of solving a reference system composed of isolated finite-sized clusters that are sometimes embedded in a uncorrelated bath media. They give access to nontrivial many body effects and can be unified within the variational principle based on the self energy functional approach \cite{P_EPJB2003}. These methods include cluster perturbation theory (CPT) \cite{SPP_PRL2000}, dynamical cluster approach (DCA) \cite{HMJK_PRB2000}, cellular dynamical mean field theory (CDMFT) \cite{KSPB_PRL2001}, variational cluster approach (VCA) \cite{PAD_PRL2003}, cluster dynamical impurity approximation (CDIA) \cite{BKSTP_EPL2009}, etc, and have been widely applied to detect the topological properties of interacting topological band insulators \cite{YXL_PRL2011,WRLH_PRB2012,LRTR_PRB2014,GMCBB_NJP2015,WFSM_PRB2016,LS_PRB2018}.

\par In this article we focus on the scheme of applying quantum cluster approaches to the determination of topological invariants in correlated Chern insulators. The prototype is the half-filled Haldane Hubbard model \cite{HZKL_PRB2011-1,HZKL_PRB2011-2,MR_PRB2013,HRP_PRB2015,ZSWZ_PRB2015,HCPP_PRL2016,VSLTHT_PRL2016,IWT_PRB2016,ASHP_PRB2016,GJMP_PRB2016,GR_NJP2018,LTTNNH_PBCM2018}, which exhibits quantum anomalous Hall effect with weak Coulomb interactions and is antiferromagnetically ordered in the strong coupling limit. Previous quantum cluster methods \cite{WFSM_PRB2016} predicted a topological Neel antiferromagnet and/or a topological nonmagnetic Mott insulator in the moderate interacting regime. These intermediate phases were claimed to be topological because their interacting bulk Chern numbers constructed from the topological Hamiltonians within the natural unit cell were shown to be nonzero. According to the generalized bulk-edge correspondence appropriate to correlated topological insulators \cite{G_PRB2011,EG_PRB2011}, there should exist gapless states or zero-energy zeros of the single particle Green's functions on open boundaries for these predicted phases. However, this self-consistency is seldom verified with an actual check in the common scheme of the quantum cluster approach community but is usually argued to hold. Unexpectedly, with close investigations on this model subject to both periodic and open boundary conditions, we find no such gapless modes nor zero-energy zeros on the edges of these intermediate phases although the bulk interacting Chern number within the natural unit cell is confirmed to be nonzero in our CPT and VCA results. We attribute this paradox to the falsity of this natural-unit-cell Chern number and elaborate that this artifact arises from the explicit breaking of the translation symmetry that encounters in most quantum cluster approaches. Moreover, we assert that the faithful interacting bulk Chern number in the framework of quantum cluster approaches must be computed in the enlarged unit cell (same as the tiling cluster), which is free of the fault caused by the explicit translation symmetry breaking and consistent with the interacting bulk-edge correspondence.

\par The rest of the paper is organized as follows. In Sec. \ref{MM}, we give a brief review on the Haldane Hubbard model and the CPT/VCA methods. In Sec. \ref{R}, we present the numerical results to elaborate the falsity of the interacting bulk Chern number calculated within the natural unit cell and verify the validity of the interacting bulk Chern number computed in the enlarged unit cell. Section \ref{SD} provides a summary and discussion.

\section{Model and Method}\label{MM}

\begin{figure}
\centering
\includegraphics[scale=1.0]{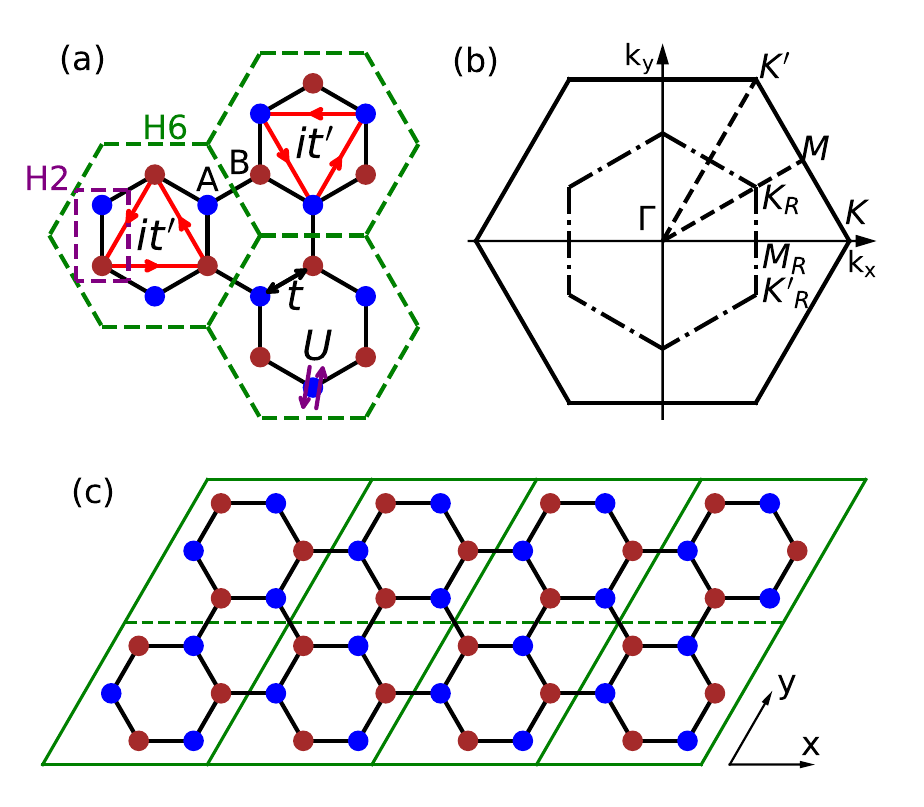}
\caption{(color online). (a) Six-site cluster tiling as enclosed by the green dashed lines on the honeycomb lattice. The natural unit cell enclosed by the purple dashed lines contains two nonequivalent sites and is denoted as H2. The enlarged unit cell denoted as H6 is equal to the six-site cluster. The hopping amplitudes $t$, $it'$ and the Hubbard interaction $U$ of the Haldane Hubbard model are also shown. (b) First Brillouin zone (FBZ) of the original lattice (black solid hexagon) and of the superlattice of the six-site cluster (black dashed hexagon). (c) An illustration of tiling the armchair ribbon used for the calculations of edge states. The superclusters (parallelogram with green solid lines) are arranged periodically along the $x$ direction. For illustration, we only plot two clusters (separated by the green dashed line) in each supercluster, while in the calculations fifteen clusters are included along the $y$ direction.}\label{lattice}
\end{figure}

\par The Haldane Hubbard model is written as $H=H_0+H_U$ as shown in Fig. \ref{lattice}(a), where $H_0$ is the spinful version of the Haldane model on the honeycomb lattice with the next-nearest-neighbor (NNN) hopping phase $\phi=\pi/2$\cite{H_PRL1988},
\begin{equation}\label{H0}
  H_0=-t\sum_{\langle ij\rangle\sigma}c^{\dagger}_{i\sigma}c_{j\sigma}-it'\sum_{\langle\langle ij\rangle\rangle\sigma}\nu_{ij}c^{\dagger}_{i\sigma}c_{j\sigma}-\mu \sum_{i\sigma}c^{\dagger}_{i\sigma}c_{i\sigma}
\end{equation}
and $H_U$ is the Hubbard interaction,
\begin{equation}\label{HU}
  H_U=U\sum_i n_{i\uparrow}n_{i\downarrow}
\end{equation}
Here, $\langle ij\rangle$ and $\langle\langle ij\rangle\rangle$ denote the nearest-neighbor and the NNN bonds, respectively. $\nu_{ij}=+1 (-1)$ if the electron makes a left (right) turn to get to the NNN site. $\mu$ is the chemical potential and is set to be $U/2$ due to the particle hole symmetry of the system. Others are in standard notation. Throughout this work, we use $t$ as the energy unit. Due to the pure imaginary NNN hopping $it'$, this model breaks the time reversal symmetry locally and there exits nonhomogeneous flux in the honeycomb lattice but the total flux through every single hexagon is zero. As a consequence, in the non-interacting limit, the ground state of the system has a nonzero Chern number and is a Chern insulator.

\par CPT \cite{SPP_PRL2000} is a quantum cluster approach accessing to the single particle Green's functions of systems with only onsite interactions. In CPT, the lattice is tiled into superlattice of clusters. Each cluster is solved exactly and the inter-cluster single particle terms are treated as perturbations. Then the single particle Green's function $G(\widetilde{\mathbf{k}},\omega)$ of the system is obtained through a RPA-like approximation:
\begin{equation}\label{SPGF}
  G^{-1}(\widetilde{\mathbf{k}},\omega)=G_0^{-1}(\omega)-V(\widetilde{\mathbf{k}})
\end{equation}
where $G_0(\omega)$ is the exact cluster single particle Green's function, $V(\widetilde{\mathbf{k}})$ is the intercluster quadratic terms, and $\widetilde{\mathbf{k}}$ belongs to the reduced first Brillouin zone (FBZ) of the superlattice. As is common in most quantum cluster approaches (except DCA), this tiling of superlattice usually partially breaks the translation symmetry, thus the single particle Green's function is not diagonal with respect to the momenta $\mathbf{k}$/$\mathbf{k}^\prime$ that belong to the FBZ of the original lattice \cite{S_arXiv2008},
\begin{equation}\label{FT}
G_{\alpha\beta}(\mathbf{k},\mathbf{k}^\prime,\omega)=\frac{1}{L}\sum_{a\in\{\alpha\},b\in\{\beta\}}
\delta(\mathbf{k}-\mathbf{k}^\prime+\mathbf{K}^\prime-\mathbf{K}) G_{ab}(\mathbf{k},\omega) \exp\left(-i\mathbf{k}\cdot\mathbf{r}_a+i\mathbf{k}^\prime\cdot\mathbf{r}_b\right)
\end{equation}
where $L$ is the number of unit cells in the cluster, $\alpha$ ($\beta$) is a composite index of site and spin in a unit cell, $\{\alpha\}$ ($\{\beta\}$) is the set of all equivalent indices of $\alpha$ ($\beta$) in the cluster with respect to the translation symmetry of the original lattice, and $\mathbf{K}$ ($\mathbf{K}^\prime$) is determined by the unique decomposition of $\mathbf{k}$ ($\mathbf{k}^\prime$): $\mathbf{k}=\widetilde{\mathbf{k}}+\mathbf{K}$ ($\mathbf{k}^\prime=\widetilde{\mathbf{k}^\prime}+\mathbf{K}^\prime$). The single particle Green's function $G_{per}(\mathbf{k},\omega)$ that respects the full translation symmetry can be approximated by the so-called periodization procedure with $\mathbf{K}^\prime=\mathbf{K}$:
\begin{equation}\label{PP}
\left[G_{per}\right]_{\alpha\beta}(\mathbf{k},\omega)=
\frac{1}{L}\sum_{a\in\{\alpha\},b\in\{\beta\}}G_{ab}(\mathbf{k},\omega)\exp\left[-i \mathbf{k}\cdot(\mathbf{r}_a-\mathbf{r}_b)\right]
\end{equation}
This approximation ignores the breaking of translation invariance caused by the different treatments given to intercluster and intracluster single particle terms. Yet it does not affect the density of states \cite{S_arXiv2008}. CPT has also been successfully applied to the studies of the single-particle spectral functions \cite{SPP_PRL2000,ST_PRL2004,KYXL_PRB2011,YXL_PRL2011,YL_PRB2012,LYXL_PRB2014}.

\par CPT is exact in the non-interacting limit but does not allow for spontaneous symmetry breaking phases. To find such phases, one would resort to VCA \cite{PAD_PRL2003}, which can be viewed as a variational version of CPT according to the self energy functional theory \cite{P_EPJB2003}. Here, the decoupled cluster with additional symmetry-breaking Weiss terms is employed as the reference system and its self-energy takes on an approximation of the original's. Then the grand potential functional can be expressed as
\begin{equation}\label{GP}
\Omega(h^\prime)=\Omega^\prime(h^\prime)-\Tr\ln\left\{1-\left[g_0^{\prime-1}(h^\prime)-g_0^{-1}\right]G^\prime_0(h^\prime)\right\}
\end{equation}
where $\Omega$ ($\Omega^\prime$) and $g_0$ ($g_0^\prime$) are the grand potential and the free propagator of the original (reference) system, respectively. $G^\prime_0$ is the exact cluster single particle Green's function of the reference system and $h^\prime$ denotes the introduced symmetry-breaking Weiss field. A symmetry-breaking phase is obtained if the grand potential $\Omega(h^\prime)$ takes a minimum at a finite value of the Weiss field. We note that the variational process is free of the periodization procedure due the trace in Eq. (\ref{GP}).

\par CPT/VCA can also calculate single particle edge states of two dimensional interacting systems. This is done by tiling clusters into a supercluster with a ribbon geometry, which is subject to open boundary condition along one direction but to periodic boundary condition along the other [e.g. see Fig. \ref{lattice}(c)]. Now the supercluster Green's function $G_0(\omega)$ in Eq. (\ref{SPGF}) is taken to be the direct sum of the single cluster Green's functions, and the couplings between individual clusters within this supercluster are treated as perturbations, i.e. they show up in the $V(\widetilde{\mathbf{k}})$ term in Eq. (\ref{SPGF}). Other procedures are exactly the same with those to compute the bulk properties.

\par In interacting many-body systems, the exact single particle Green's function $\mathcal{G}(\mathbf{k},\omega)$ is a useful tool to investigate the topological properties \cite{V2003,G_PRB2011,WZ_PRX2012}. Actually, the topological Hamiltonian \cite{WY_JPCM2013}, which is defined as the minus inverse of the zero-frequency single particle Green's function, i.e.
\begin{equation}\label{TH}
H_{topo}(\mathbf{k})=\left.-\mathcal{G}^{-1}(\mathbf{k},\omega)\right|_{\omega=0}
\end{equation}
is sufficient to encode all the topological information \cite{WZ_PRX2012}. For example, the interacting Chern number can be constructed via the Kubo formula \cite{TKNN_PRL1982}
\begin{equation}\label{KF}
C=i\sum_{E_m<\mu<E_n} \int\frac{d^2\mathbf{k}}{(2\pi)^2} \frac{\langle m|v_x|n\rangle\langle n|v_y|m|\rangle-\langle m|v_y|n\rangle\langle n|v_x|m|\rangle}{(E_m-E_n)^2}
\end{equation}
in terms of the topological Hamiltonian's eigenvalues $E_i$ and eigenvectors $|i\rangle$ ($i=m,n$). Here $v_\alpha=\frac{\partial H_{topo}}{\partial k_\alpha}$ ($\alpha=x,y$) is the velocity operator. In the framework of quantum cluster approaches, the calculation of the interacting Chern number is performed by approximating the $\mathcal{G}(\mathbf{k},\omega)$ in Eq. (\ref{TH}) by the $G_{per}(\mathbf{k},\omega)$ in Eq. (\ref{PP}).

\section{Results}\label{R}

\par Our results are obtained by use of CPT/VCA on the 6-site-cluster-tiled superlattice (the superlattice is composed of 100$\times$100 clusters) as illustrated in Fig. \ref{lattice}(a). This cluster not only preserves the point-group symmetry of the Hamiltonian but also has zero net flux through it, which is essential for the Haldane Hubbard model. So, it is the best choice considering the available computation capacity.

\subsection{Phase diagram}

\begin{figure}
\centering
\includegraphics[scale=0.45]{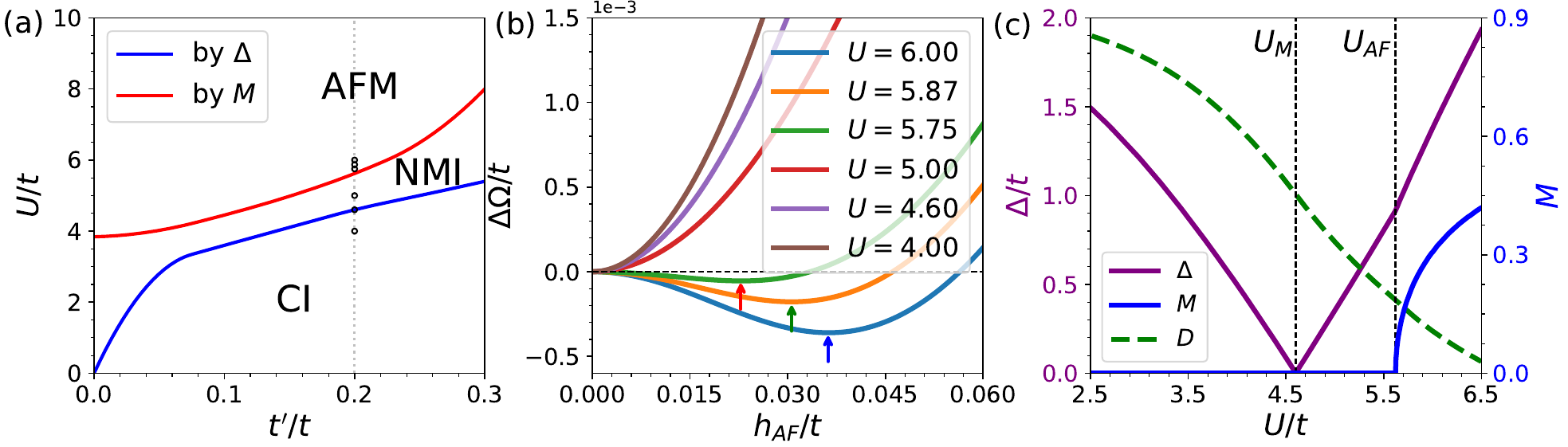}
\caption{(color online). (a) Groundstate phase diagram of the Haldane Hubbard model. CI, NMI, and AFM denote Chern insulator, nonmagnetic Mott insulator and antiferromagnetic Mott insulator, respectively. The dotted line at $t'=0.2$ marks the parameters used in Fig. \ref{phase}(c), Fig. \ref{h2chernnumber}(b) and Fig. \ref{h6chernnumber}(b). The circles mark the parameters used in Fig. \ref{phase}(b), Fig. \ref{h2spectra}, Fig. \ref{edge}, and Fig. \ref{h6spectra}. (b) $\Delta\Omega$ as a function of $h_{AF}$ for $t'=0.2$. Arrows indicate the positions of the minima, where magnetic solutions are allowed. (c) The single particle gap $\Delta$ (purple solid line), the antiferromagnetic moment $M$ (blue solid line) and the double occupancy $D$ (green dashed line) as a function of $U$ for $t'=0.2$. $U_M=4.60$ and $U_{AF}=5.62$ represent the Mott transition point and the paramagnetic-antiferromagnetic transition point, respectively. We note that the double occupancy is rescaled linearly in order to plot it in the same figure.} \label{phase}
\end{figure}

\par Before the discussion of the interacting Chern number $C$, we first present the phase diagram determined by the single particle gap $\Delta$, the antiferromagnetic (AF) moment $M=\frac{1}{N}\sum_i(-1)^{\eta_i}\langle c^\dagger_{i\alpha}\sigma^z_{\alpha\beta}c_{i\beta}\rangle$ and the double occupancy of electrons $D=\frac{1}{N}\sum_i\langle n_{i\uparrow}n_{i\downarrow}\rangle$, as shown in Fig. \ref{phase}(a). Here $\eta_i=0$ or $1$ when $i\in A$ or $B$ and $N$ is the number of lattice sites. There are three distinct phases separated by two phase boundaries. The single particle gap keeps nonzero except on the transition line from the CI to the nonmagnetic Mott insulator (NMI) while the AF moment is nonzero only for the antiferromagnetic Mott insulator (AFM).

\par Take $t'=0.2$ as an example, let's have a close look at these quantities. $\Delta$, $M$ and $D$ are computed at the minima of the grand potential $\Omega$ with respect to the Weiss field corresponding to the AF order: $H_{AF}=h_{AF}\sum_i(-1)^{\eta_i}c^\dagger_{i\alpha}\sigma^z_{\alpha\beta}c_{i\beta}$. In Fig. \ref{phase}(b), $\Delta\Omega=\Omega(h_{AF})-\Omega(h_{AF}=0)$ at the parameters indicated by the circles in Fig. \ref{phase}(a) are shown. It is clear that for small Hubbard $U$, $\Delta\Omega$ takes its minimum when $h_{AF}=0$, indicating that the system is exempt from magnetic order, while for large $U$, $\Delta\Omega$ reaches its minimum when $h_{AF}$ is nonzero, meaning that the AF moment has already formed. In Fig. \ref{phase}(c), $\Delta$, $M$ and $D$ are plotted for a range of $U$s. With increasing $U$, $\Delta$ decreases at first, gets to zero at $U_M$, and then increases linearly, but encounters a change of the slope at $U_{AF}$. A rapid decrease of $D$ occurs around $U_M$, specifically the inflection point of $D$ coincides with $U_M$. Combining this with the linear dependency of the reopened gap with $U$, the transition at $U_M$ is identified as the Mott transition. $M$ develops continuously at $U_{AF}$. Therefore, the transition at $U_{AF}$ is the paramagnetic-antiferromagnetic transition.

\subsection{Paradox between the natural-unit-cell bulk Chern number and the edge states/zeros}\label{Paradox}

\begin{figure}
\centering
\includegraphics[scale=0.6]{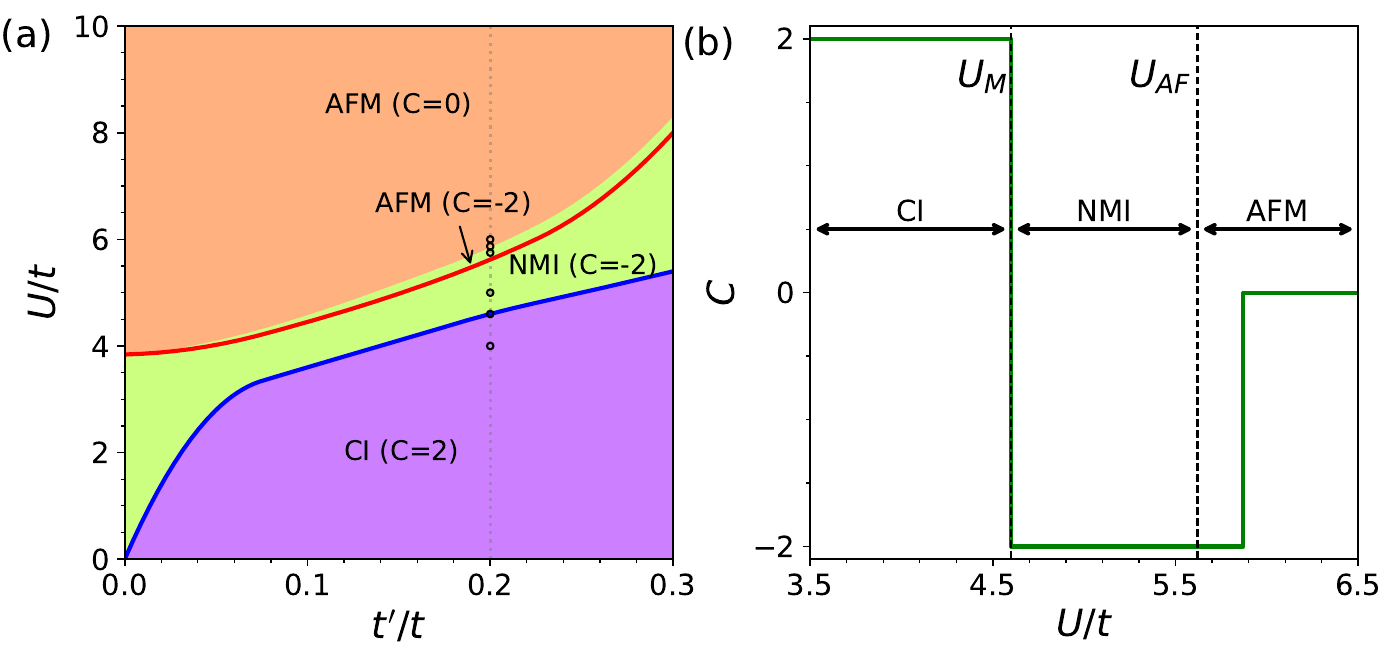}
\caption{(color online). Interacting Chern number $C$ calculated within the natural unit cell (H2 in Fig. \ref{lattice}) (a) in the $U$-$t'$ plane (2 for magenta, -2 for green and 0 for orange), and (b) as a function of $U$ with $t'=0.2$. The AFM phase contains a narrow region with a nonzero Chern number near the AFM-NMI phase boundary. The dotted line and the circles are the same as those in Fig. \ref{phase}(a).}\label{h2chernnumber}
\end{figure}

\par Now let's turn to the interacting Chern number $C$ of the model, which is commonly computed in terms of Eqs. (\ref{TH}-\ref{KF}) within the natural unit cell [H2 in Fig. \ref{lattice}(a)], as shown in Fig. \ref{h2chernnumber}(a). It is observed that the Chern number is $2$ for CI and $-2$ for NMI, while for the AFM phase, it is $-2$ near the NMI-AFM phase boundary and $0$ deep in the AFM phase. To see this more clearly, in Fig. \ref{h2chernnumber}(b), we plot $C$ as a function of $U$ with $t^\prime=0.2$. Apparently, the Chern number changes its sign across $U=U_M$ and survives a noticeable range of $U$ after the AF order appears, and eventually alters its value from $-2$ to $0$ at a value of $U>U_{AF}$. Therefore, it seems that there exist two successive topological phase transitions.

\begin{figure}
\centering
\includegraphics[scale=0.45]{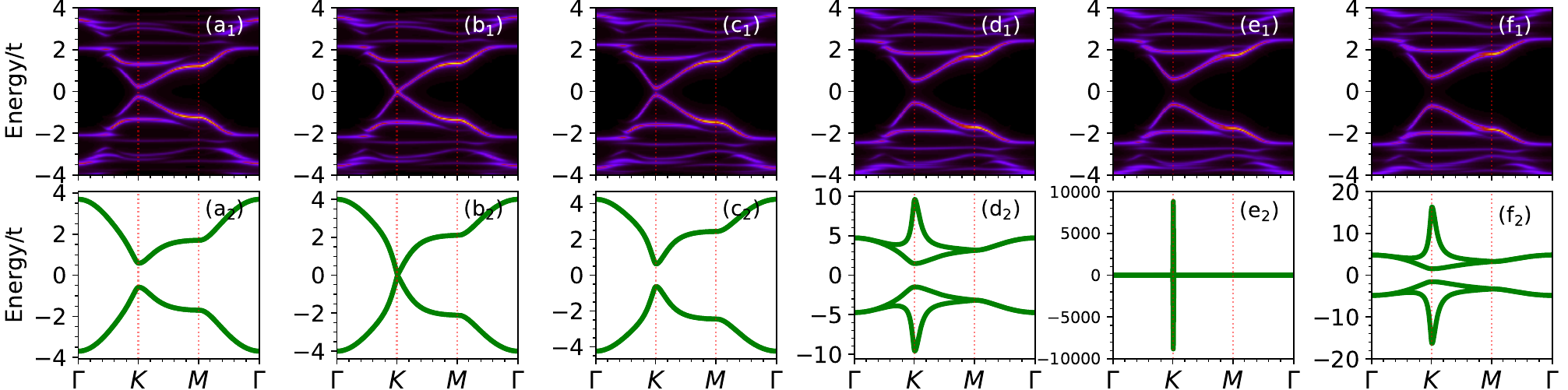}
\caption{(color online). Intensity plot of the single particle spectra (top) and topological Hamiltonian's spectra (bottom) along the high symmetry path in the FBZ calculated with (a) $U=4.00$, $h_{AF}=0$, (b) $U=4.60$, $h_{AF}=0$, (c) $U=5.00$, $h_{AF}=0$, (d) $U=5.75$, $h_{AF}=0.0228$, (e) $U=5.87$, $h_{AF}=0.0306$, and (f) $U=6.00$, $h_{AF}=0.0362$. The values of $h_{AF}$ are determined by the locations of $\Delta\Omega$s' minima in Fig. \ref{phase}(b). The NNN hopping is fixed at $t'=0.2$. (b) corresponds to the first change of the natural-unit-cell Chern number $C$ in Fig. \ref{h2chernnumber}(b) and (e) to the second.}\label{h2spectra}
\end{figure}

\par It is established that in the non-interacting limit, topological phase transitions cannot occur without closing the single particle gap. This also accounts for what happens at the first transition ($U=U_M$) when the interaction is not strong enough to induce an AF order in the model studied here. In the interacting regime, there are two more scenarios of topological transitions. One is the first-order transition across which the local order parameters change discontinuously \cite{BTS_PRB2013,ZSWZ_PRB2015,ABCTS_PRL2015}, the other is the zeros-driven transition where the change of the topological invariants is due to the emergence of zero-energy zeros of the single-particle Green's functions \cite{G_PRB2011,BTLLZ_PRB2012,SK_PRB2018}. Both scenarios can avoid the closing of the single particle gap. From Fig. \ref{phase}(c), we can see that at the ``second topological transition" ($U>U_{AF}$), the single particle gap $\Delta$ does not close and the AF moment $M$ is continuous with $U$. Thus this transition does not belong to the first two scenarios. Then does it belong to the last scenario? To see this, in Fig. \ref{h2spectra}, we plot the single particle spectral function $A(\mathbf{k},\omega)=-2\,{\rm Im} G(\mathbf{k},\omega+i0^+)/\pi$ and the topological Hamiltonian's spectra along a high symmetry path in the FBZ at the circled points indicated in Fig. \ref{h2chernnumber}(a). With increasing $U$, the single particle spectra and the topological Hamiltonian's spectra show a quite similar evolution before the AF moment forms [see Figs. \ref{h2spectra}(a)-(c)], namely their gaps both close at $K$/$K'$ at the first transition ($U=U_{M}$) and reopen after that. However, their behaviors deviate from each other essentially after the formation of the AF moment [see Figs. \ref{h2spectra}(d)-(f)]. Now, the single particle spectra exhibit nearly no change except that the gap increases linearly. While, the topological Hamiltonian's spectra quickly develop a divergence at $K$/$K'$ when one approaches the ``second topological transition" ($U>U_{AF}$).

\begin{figure}
\centering
\includegraphics[scale=0.4]{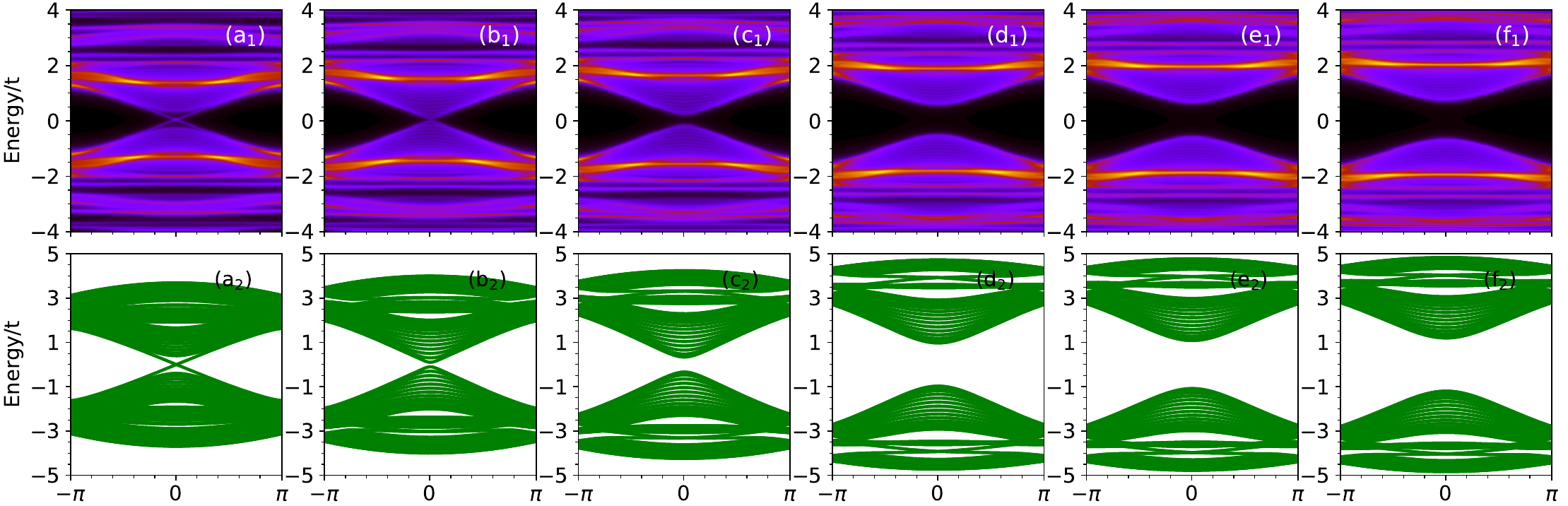}
\caption{(color online). Intensity plot of the single particle spectra (top) and topological Hamiltonian's spectra (bottom) on the armchair ribbon illustrated in Fig. \ref{lattice}(c) calculated with the same parameters in Fig. \ref{h2spectra}.}\label{edge}
\end{figure}

\par A divergence of the spectra of the topological Hamiltonian corresponds to the emergence of a zero-energy zero of the single particle Green's functions. Now it seems to lead us to a conclusion that the ``second topological transition" at $U>U_{AF}$ is a concrete realization in a microscopic model of the last scenario, as was similarly claimed in Ref. \cite{WFSM_PRB2016}. However, according to the generalized bulk-edge correspondence \cite{G_PRB2011,EG_PRB2011}, a topological phase with a nonzero interacting bulk Chern number must assume gapless single particle states or zero-energy zeros of the single particle Green's function on open boundaries. To check this, in Fig. \ref{edge}, we plot the single particle spectral and the topological Hamiltonian's spectra with the same parameters as those in Fig. \ref{h2spectra} on an armchair ribbon as illustrated in Fig. \ref{lattice}(c). In order to completely tile the stripe, a 6-site cluster is used. In our calculations, fifteen clusters are included in the $y$ direction to form a supercluster and the superclusters are arranged periodically in the $x$ direction. For illustration, we only show schematically two clusters in the $y$ direction in Fig. \ref{lattice}(c). It is clear that before the first transition at $U=U_M$, for both the single particle spectra and the topological Hamiltonian's spectra, there always exist gapless edge modes [see Fig. \ref{edge}(a)]. However, after this transition, we cannot observe any gapless edge state or any zero-energy zero of the single particle Green's function [see Figs. \ref{edge}(c)-(f)], even for those with a nonzero natural-unit-cell bulk Chern number [see Figs. \ref{edge}(c)-(d)]. Therefore, a discrepancy between the natural-unit-cell bulk Chern number and the edge states/zeros exists here.

\subsection{Falsity of the natural-unit-cell Chern number and validity of the enlarged-unit-cell Chern number}

\begin{figure}
\centering
\includegraphics[scale=0.6]{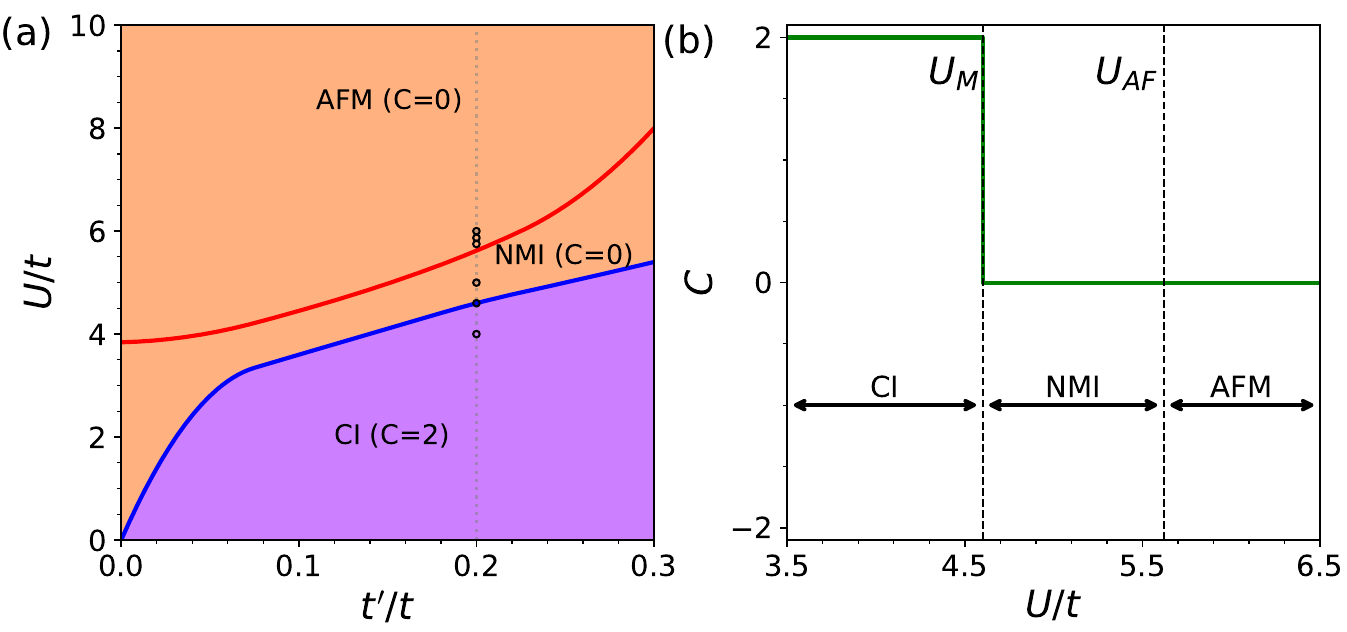}
\caption{(color online). Interacting Chern number $C$ calculated within the enlarged unit cell (H6 in Fig. \ref{lattice}) (a) in the $U$-$t'$ plane (2 for magenta, and 0 for orange), and (b) as a function of $U$ with $t'=0.2$. Both the NMI phase and the AFM phase acquire zero Chern number. The dotted line and the circles are the same as those in Fig. \ref{phase}(a).}\label{h6chernnumber}
\end{figure}

\par To resolve the above paradox, let's first enumerate the approximations we have used in our quantum cluster calculations. Firstly, the self-energy of the real system is replaced by that of a reference system. Secondly, the single particle Green's functions in the original Brillouin zone are approximated by a periodization procedure as indicated by Eqs. (\ref{FT}) and (\ref{PP}). When we study the model here with both periodic and open boundary conditions, their reference systems are the same, but the periodization procedure is needed for the periodic boundary condition while it is not for the open boundary condition. This periodization procedure aims to restore the explicitly broken translation symmetry of the original lattice but now is suspectable to be responsible for this paradox.

\begin{figure}
\centering
\includegraphics[scale=0.45]{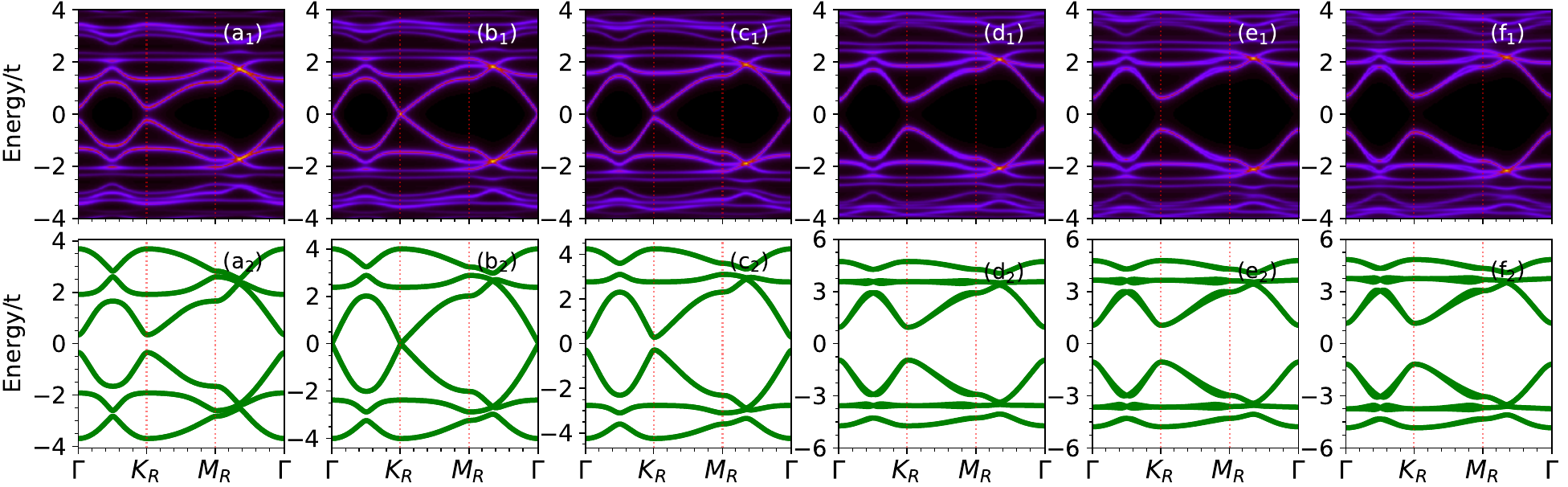}
\caption{(color online). Intensity plot of the single particle spectra (top) and topological Hamiltonian's spectra (bottom) along the high symmetry path in the reduced FBZ calculated with the same parameters in Figs. \ref{h2spectra}-\ref{edge}.}\label{h6spectra}
\end{figure}

\par To check this possibility, we repeat the computation of the Chern number by our quantum cluster approaches with the enlarged six-site unit cell [labeled as H6 in Fig. \ref{lattice}(a)], which is simply the tiling cluster of the superlattice and is thus free of the periodization procedure. The results presented in Fig. \ref{h6chernnumber} show that the enlarged-unit-cell Chern number vanishes after the Mott transition point $U_M$. Therefore, this calculation indicates that there are no topological phase in the moderate interacting regime and hence no the ``second topological transition''. Now the interacting bulk-edge correspondence is satisfied. In Fig. \ref{h6spectra}, we also plot the single particle spectra and the topological Hamiltonian's spectra in this case with the same parameters as those in Fig. \ref{h2spectra}. It can be seen that although both spectra still close their gaps at $U_M$ [see Fig. \ref{h2spectra}(b) and Fig. \ref{h6spectra}(b)], the divergence of the topological Hamiltonian's spectra does not emerge after the AF moment develops [see Fig. \ref{h2spectra}(e) and Fig. \ref{h6spectra}(e)]. Clearly, all the characteristic features of the nontrivial topology indicated by the bulk Chern number and the topological Hamiltonian's spectra calculated within the natural unit cell just disappear in our enlarged six-site unit cell calculations.

\par Thus, the enlarged-unit-cell Chern number is physically reasonable compared to the natural-unit-cell Chern number, not only because of its avoidance of the approximation caused by the periodization procedure but also due to its consistency between the bulk topological invariant and the edge states or zero-energy zeros of the single particle Green's function. The above point is further supported by the following physical arguments. Assuming that the original system is disturbed by a small local perturbation $\Delta H$ such that the translation symmetry is broken and the new unit cell is exactly the six-site cluster, a real topological phase with a truly nonzero Chern number should be stable against such small perturbations. It means that the new Chern number of the disturbed phase should remain unchanged. Actually, the enlarged-unit-cell Chern number is identical to this perturbed Chern number in the $\Delta H=0+$ limit. We note that the natural-unit-cell Chern number and enlarged-unit-cell Chern number obtained in the weak interaction regime before the Mott transition are the same, meaning that the CI phase here is real topological. While, the nontriviality of the natural-unit-cell Chern number of the intermediate phases shown in Sec. \ref{Paradox} and Ref. \cite{WFSM_PRB2016} is just an artifact of the breaking of the translation symmetry which is hard coded in the periodization procedure of quantum cluster approaches.

\section{Summary and discussion}\label{SD}

\par In summary, with investigations on the Haldane Hubbard model as the prototype, we have uncovered the unexpected possible falsity of the natural-unit-cell interacting Chern number computed by quantum cluster approaches in correlated Chern insulators. This falsity is deeply related to the explicit breaking of the translation symmetry which is universal for most quantum cluster approaches. We assert that the faithful interacting bulk topological invariant in the framework of quantum cluster approaches must be computed in the enlarged unit cell, which is free of the fault caused by the explicit translation symmetry breaking and consistent with the interacting bulk-edge correspondence.

\par We want to remark that the actual phase diagram of the Haldane Hubbard model in the moderate interacting regime has been highly debated over the past several years, namely different methods predicted quite different results \cite{HZKL_PRB2011-1,HZKL_PRB2011-2,MR_PRB2013,HRP_PRB2015,ZSWZ_PRB2015,WFSM_PRB2016,HCPP_PRL2016,VSLTHT_PRL2016,IWT_PRB2016,ASHP_PRB2016,GJMP_PRB2016,GR_NJP2018,LTTNNH_PBCM2018}. In particular, even whether the intermediate phases exist has not come to an agreement, let alone their physical properties. Due to the limited power of quantum cluster approaches, we have no intention to settle down this difficult problem in this paper. It is only safe to claim that in the framework of CPT/VCA, there exist an intermediate phase (NMI), which is not topological in the sense that it has no nonzero interacting bulk Chern number nor gapless states or zero-energy zeros of the single particle Green's function on edges. Yet, this claim cannot rule out the possibility of the exotic nonmagnetic phases with topological fractionalized excitations \cite{HZKL_PRB2011-2,MR_PRB2013,HRP_PRB2015,HCPP_PRL2016} because it is not guaranteed that the nontrivial topology of the fractionalized modes must manifest itself in the single particle Green's functions. It is also possible that the NMI phase is just an artifact of the CPT/VCA method. Indeed, as can be seen in Fig. \ref{phase}(a), the NMI phase connects to the well known fictitious ``spin liquid'' phase in the simple honeycomb Hubbard model with $t'=0$ \cite{MLWAM_N2010,SOY_SR2012,AH_PRX2013,HD_PRL2013}, which arises from the nonequivalent treatment of the intracluster hoppings and the intercluster ones of the six-site tiling\cite{LRTR_PRB2014}. Yet the experiences of the case with $t'=0$ might not be definitely generalized to that with a finite $t'$. In view of the effective spin model obtained in the large-$U$ expansion, a nonzero $t'$ term introduces next nearest neighbor spin interactions and ring-exchange spin interactions, which could frustrate magnetic order and stabilize exotic quantum disordered phases \cite{HRP_PRB2015,HCPP_PRL2016}. Besides, charge fluctuations are known to be possible to suppress magnetic order and favor nonmagnetic phases in the moderate interacting regime \cite{SS_PRL2008,YKK_PRL2009,RLRT_PRL2015}. Further sophisticated numerical analyses, such as the large scale density matrix renormalization group simulations, are needed to determine the phases in the moderate interacting regime. We note that the noninteracting spinless Haldane model has been realized in recent cold atom experiments using the shaking lattice technique \cite{JMDRLUGE_N2014}. We hope that our claim can be tested in future experiments.

\par We emphasize, whether the NMI phase exists is in fact not directly related to the reported fault here caused by the standard periodization procedure that leads to a false interacting Chern number. As mentioned in Sec. \ref{MM}, the density of states and the grand potential are not affected by the approximation employed by the periodization procedure, thus, the single particle gap $\Delta$ and the antiferromagnetic moment $M$ are irrelevant to this approximation. Note that $\Delta$ and $M$ completely determine the phase diagram shown in Fig. \ref{phase}(a). Therefore, the cause that accounts for a false interacting topological Chern number is different from the one that may result in a possible fictitious NMI. Thus, the validity of the modified computing scheme of applying quantum cluster approaches to the calculation of interacting topological invariant is independent of the existence of the ``NMI'' phase. In fact, it applies to all insulating phases that could either be nonmagnetic or host magnetic orders. Indeed, based on the same computing scheme, the whole AFM phase is shown to be non-topological, which invalidates the claim in a previous quantum cluster study based on the false natural-unit-cell Chern number that there exists a topological region in the AFM phase in the moderate interacting regime.

\section*{Acknowledgement}
\par We acknowledge helpful discussions with S. -L. Yu, Z. Wang, J. -G. Liu and W. Wang. This work was supported by the National Natural Science Foundation of China (No.11774152, No.11704341) and National Key Projects for Research and Development of China (Grant No. 2016YFA0300401).

\section*{References}
\bibliography{ref}

\end{document}